\documentclass[10pt,twocolumn]{article} 

\usepackage{a4wide}
\usepackage[utf8]{inputenc}

\usepackage{amsmath,amssymb,graphicx}
\usepackage{bm}

\usepackage{multirow}

\newcommand{\inv}[1]{\frac{1}{#1}}
\newcommand{\comb}{\text{comb}}
\newcommand{\N}{\mathbb{N}}
\newcommand{\R}{\mathbb{R}}
\newcommand{\Z}{\mathbb{Z}}
\newcommand{\ud}{\mathrm{ d }}
\newcommand{\half}{\frac{1}{2}}

\newcommand{\unit}[1]{\ensuremath{\,\, \mathrm{#1}}}

\makeatletter
{\begin{center}\begin{minipage}{\textwidth}}%
{\end{minipage}\end{center}\vspace{2mm}}%
\newskip\abstlineskip%
\makeatother
\begin{document}

\title{On the Optical Role of Randomness for Structured Surfaces}

\author{Villads Egede Johansen\\
\\
Department of Mechanical Engineering, \\
Technical University of Denmark,\\
2800 Kgs. Lyngby, Denmark\\
\\
vejo@mek.dtu.dk\\
}

\maketitle
\thispagestyle{empty}

\begin{abstract}
It has been known for years how random height variations of a repeated nano-scale structure can give rise to smooth angular color variations instead of the well-known diffraction pattern experienced if no randomization is present. However, until now there has not been published any papers giving an in-depth mathematical explanation on the mechanisms behind and how to design the randomness for a given application. This paper presents a mathematical framework for analyzing these random variations -- rigorously as well as intuitively.
\end{abstract}

\section{Introduction}
In the 17th century, Robert Hooke discovered how dielectric structures with size features comparable to the wavelength of light were an important part of the color and appearance that was found in certain animals he studied under a microscope  \cite{Hooke1665}. However, the theory of light was not well developed at the time, there were no computers, and the quality of microscopes was not good either so the field of structural colors remained rather untouched. In recent decades, the invention of electron microscopes and the computer and the wave theory of light in the last century have made it possible to do more in-depth investigation of this field that has more than 500 million years of history in nature \cite{Parker2000}.

Controlling light reflection by interaction with structures is crucial for many applications. Retro-reflectors, aluminum foil, solar cells and security holograms are just few examples encountered frequently in everyday life that would not work well without. These examples also show that there are many motivations for controlling color appearance of an object besides the visual appearance, and that improving the understanding of light's interaction with surfaces can improve a wide range of engineering applications and possibly initiate new inventions.

Many important contributions to the understanding of structural colors have been discovered during the study of the nanostructure of the wing of the Morpho rhetenor butterfly. The results range from what is presented in \cite{Vukusic1999} and up until present day, where it is possible to make a computer model of the wing's color appearance and reflection based on measurements of the structure of the wing \cite{Okada2012}. On the Morpho type in general, excellent works have been published dating longer back, see e.g. \cite{Mason1927,Lippert1959}. The reasons for why especially this structure has become so central in the analysis of structural colors are probably (1) the fact that its structure is more or less invariant along one axis, leaving it possible to simulate only a cross section of the model; (2) and the rather simple shape of the structure, making it possible at an early stage to obtain good results just by analyzing it as a multilayer structure and then elaborate more and more on that model.

One of the properties which still needs investigation is how to model the random displacements of the individual, repetitive structures present in the Morpho rhetenor's wing, see Figure 3.45 in \cite{Parker2012} and the following description in \cite{Kinoshita2012}. By numerical as well as practical experiments it has been shown how these random displacements of a periodic structure seem to smooth out the otherwise strong diffractive effects which are expected from reflections of periodic structures. A good mathematical explanation for this -- that can also be used in a wider setting -- is missing. In this paper we will present a mathematical framework for analysing repeated structures with (or without) random translations of the elements. As a benchmark, previously published results on the random behaviour related to the Morpho rhetenor butterfly will be used, but the framework has a much broader aim than this: it should make it possible to design structures with new color effects, and also it should help giving a better intuitive understanding of the influence of different kinds of randomization of structures (e.g. in-plane vs. height displacements). The theory treated here will focus on the visible light, but can be applied to all parts of the electromagnetic spectrum.

The rest of the paper is organized as follows. Section \ref{sec:observations} motivates the work in this paper by showing examples in the literature where it can be used, Section \ref{sec:framework} deals with the mathematical background needed to analyze the random effects, Section \ref{sec:analysis} applies this theory in a general setting to results obtained in earlier studies and shows how these could have been predicted by this framework, Section \ref{sec:colors} gives some examples on what influence the randomness would have on the color appearance of a surface for some specific cases, and finally Section \ref{sec:conclusion} concludes on the the presented results.

\section{Observations of randomness in the literature} \label{sec:observations}
In this paper we will focus on surfaces comprised of repeated unit structures in the $x,y$-plane with some per unit height displacement in the $z$-direction. See Figure \ref{fig:cellstructure} for an example of a repeated structure. This is because structures fitting to this description are found many places in the literature of optics and within many different fields. Partly where the translations of the copied structures are deterministic and also where the description contains random parameters. To motivate the introduction of the following framework, some of the publications relying on one unit structure repeated throughout a domain are listed:

\begin{itemize}
 \item Firstly, several publications exist considering the influence of randomness on binary gratings: in \cite{Licinio1999} an experimental study of in-plane randomness is conducted with results that can be seen as a special case of the following framework; and in \cite{Rico-garcia2009} some binary height variations of gratings are studied that also can be considered a special case of the framework.
 \item Secondly, phase gratings using the properties of randomness for concrete products like e.g. a surface giving \emph{controlled angular redirection of light} for windows to improve indoor lighting environment has been presented in \cite{Buss2013}. Here the effect of in-plane randomness is observed, but no mathematical explanation is given.
 \item Thirdly, designing random disorder is also seen in photovoltaic solar cell applications as presented in e.g. \cite{Pratesi2013}. In this article the following framework could have been used to investigate the effect of different randomization algorithms and parameters related to that before utilizing full wave simulations for a detailed study.
\end{itemize}
For testing the usefulness of the framework, studies of the nanostructure of the Morpho butterfly's wing will be used, since it has undergone many studies the last decades with focus on different aspects of randomness. The works to be used are:
\begin{itemize}
 \item \emph{Detailed electromagnetic simulation for the structural color of butterfly wings}, \cite{Lee2009}, which shows numerically how the far-field response of one lit Morpho butterfly ridge (which is the ``unit structure'' of this butterfly's wing) almost corresponds to the response of many random height translated elemenents;
 \item \emph{Reproduction, Mass-production, and Control of the Morpho-butterfly's Blue}, \cite{Saito2009}, which shows by experiment how a binary random pattern with a structure on top can be used to generate a smooth color effect compared to no randomization;
 \item \emph{Numerical Analysis on the Optical Role of Nano-Randomness on the Morpho Butterfly's Scale}, \cite{Saito2011}, which conducts several numerical experiments with different kinds of randomization parameters to investigate the effect of these, and;
 \item \emph{Detailed simulation of structural color generation inspired by the Morpho butterfly}, \cite{Steindorfer2012}, which -- among other numerical experiments -- contains an analysis of the effect of different maximum heights chosen for randomization.
\end{itemize}
In this paper we confirm these observations by mathematical derivations of a simple framework and numerical experiments performed using that framework.

\section{Framework for analyzing random translations} \label{sec:framework}
In this section a framework is presented which can be used for analysis of (random) translations of structures. It is presented for a 3D general case, even though the examples later on will be two-dimensional. The analysis is performed for time harmonic waves using the time factor $e^{j\omega t}$.

\subsection{Huygens' principle}
\begin{figure}
  \centering
 \includegraphics[width=\columnwidth]{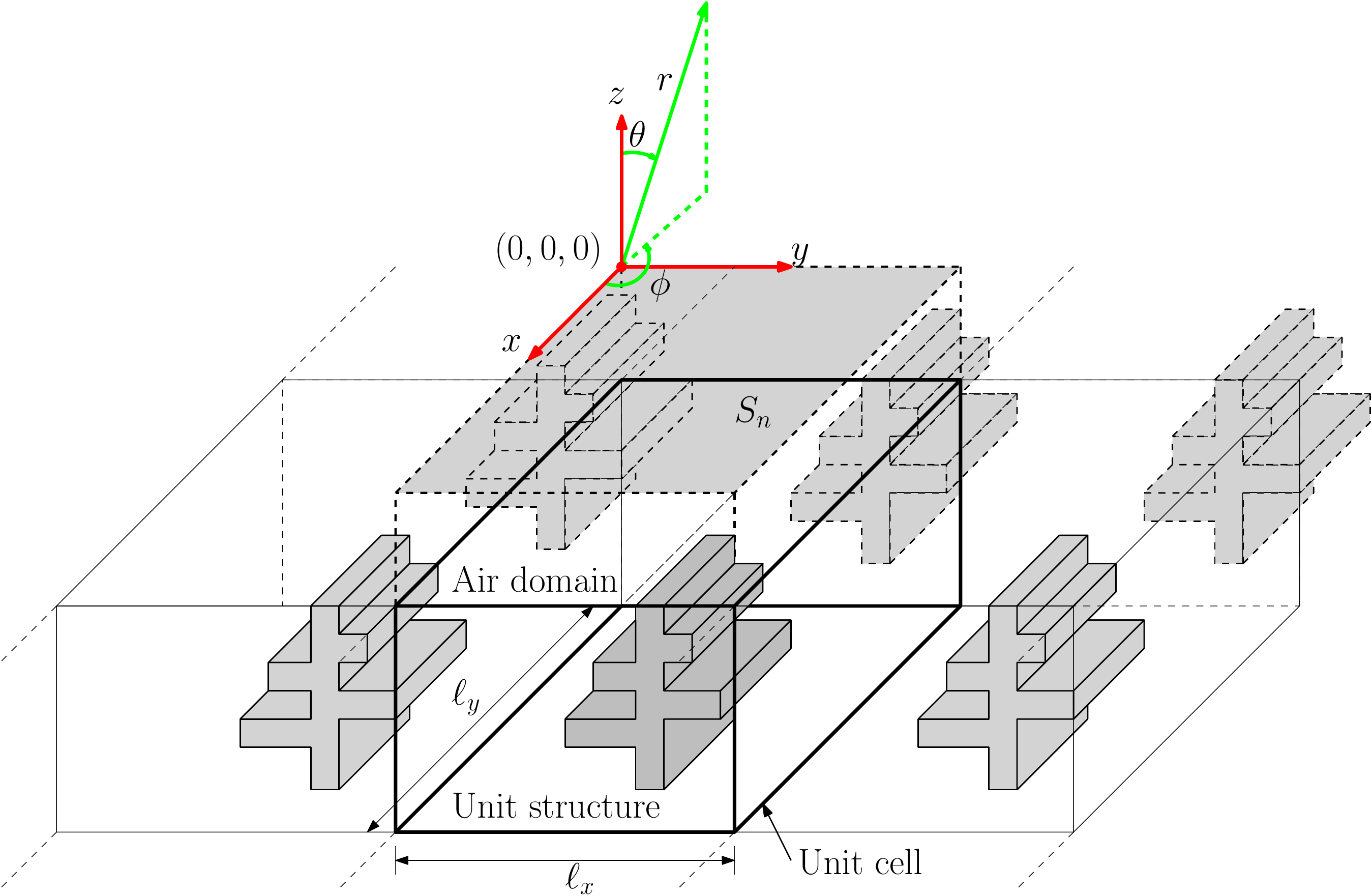}
  \caption{Some structure divided into cells with an air domain above.}
  \label{fig:cellstructure}
\end{figure}

Consider an electromagnetic structure with its volumes divided into different cells -- which in this article will be referred to as units when all cells contain the same structure -- as seen in Figure \ref{fig:cellstructure}. By Huygens' principle and the image principle, \cite{Balanis2012}, it is possible to calculate the (near-field as well as) far-field contribution in the upper hemisphere from any structure by considering the so-called equivalent surface currents, $\bm J_e$, calculated on an infinite plane above the structure. An example of this could be the structure seen in Figure \ref{fig:cellstructure} if the plane $S_n$ is extended to infinity. Due to the linearity of Maxwell's equations, the plane can be split into parts with each their far-field contribution, and by summing them we still arrive at the same total far-field contribution.

Now, by defining a plane above the structure in Figure \ref{fig:cellstructure} and splitting it into parts such that each part follows the projection of the cell on the plane -- like $S_n$ in Figure \ref{fig:cellstructure} -- the magnetic far-field contribution for the $\bm H$-field, $\bm H^f$ can be found as (see e.g. Chapter 6 in \cite{Balanis2012})
\begin{align}
 \bm H^f (\theta,\phi)= \sum _{n\in \N} \bm H^f_n(\theta,\phi),
\end{align}
where
\begin{align}
\bm H^f _n (\theta,\phi) =  - jk \frac{e^{-jkr}}{4\pi r} \bm {\hat r} \times \int_{S_n} \bm J_e e^{jk \bm {\hat r}(\theta,\phi) \cdot \bm r_o} \ud S_n, \label{eqn:Hf}
\end{align}
where $k=2\pi/\lambda$ is the wavenumber, the $S_n$'s ($n \in \N$) are all unique parts making up the total surface $S$, $r$ is the distance from an arbitrarily located origin on $S_n$ to the evaluation point (since $r$ is used in places where phase information is not important, it is assumed constant), $\bm {\hat r} = (\sin \theta \cos \phi, \sin \theta \sin \phi, \cos \theta)$ is the direction towards the far-field evaluation point, and $\bm r_0=(x,y,z)\in S_n$ is the position vector (as measured from the origin) to a point on $S_n$. The variables $x,y,z,\theta,\phi$ are all defined as in  Figure \ref{fig:cellstructure}. 

In far-field and free space, the radiated wave locally approaches a TEM (Transverse ElectroMagnetic) wave, and we can therefore make use of the relation between the electric and magnetic field due to this behavior,
\begin{align}
 \bm E^f = \eta \bm H^f \times \bm {\hat{r}},
\end{align}
where $\eta \approx 377 \unit{\Omega}$ is the free-space impedance, to calculate the irradiance $E$ -- which equals the magnitude of Poynting's vector -- as
\begin{align}
E &= \half\left|  \bm E^f \times \overline{\bm H^f} \right| = \half \eta |\bm H^f|^2 = \half \eta \bigg| \sum _{n \in \N} \bm H^f_n \bigg|^2, \label{eqn:irradiance}
\end{align}
which will be used later on in the analysis.

\subsection{Translation of lit structures} \label{sec:translation}
\begin{figure}
\begin{center}
 \includegraphics[width=\columnwidth]{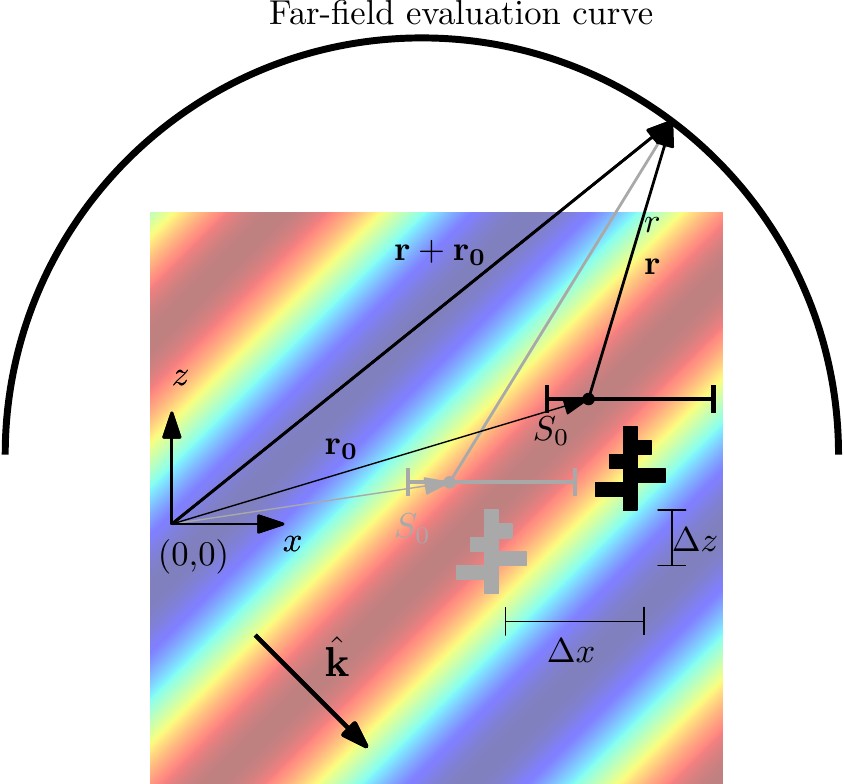}
 \caption{Geometry of the translation of a structure lit by a plane electromagnetic wave.}
 \label{fig:translation_example}
 \end{center}
\end{figure}

We want to manipulate \eqref{eqn:irradiance} such that it includes (random) height/length translations of structures in all three spatial directions, $\Delta \bm r = (\Delta x,\Delta y, \Delta z)$, see Figure \ref{fig:translation_example} for a planar example. Considering one of these structures, it is lit by a plane wave with propagation direction $\hat {\bm k}$ as also indicated in the figure. This means, first of all, that by translating it, we introduce a phase lag, $\Delta p$, on the phase that the incoming wave meets the structure with, which can be described as
\begin{align}
 \Delta p = -\hat {\bm k} \cdot \Delta \bm r/\lambda.
\end{align}
In relation to the evaluation of the far field there is also a difference, since the vector $\bm r _0$ has changed so that it now is represented by
\begin{align}
 \bm r _ 0 ^{new} = \bm r _0 + \Delta \bm r,
\end{align}
and this influences the far-field transformation in \eqref{eqn:Hf}. We do not need to update the distance, $r$, since the change in contribution to the distance is negligible -- which is in agreement with the assumptions made in the derivation of far-field radiation in the first place.

Modifying \eqref{eqn:Hf} then gives
\begin{align}
\nonumber \bm H^{f,t} _n (\theta,\phi) &=  - e^{-j2\pi \hat{\bm k} \cdot \Delta \bm r / \lambda} jk \frac{e^{-jkr}}{4\pi r} \bm {\hat r} \times \\
\nonumber				      &\qquad \int_{S_n} \bm J_e e^{jk \bm {\hat r}(\theta,\phi) \cdot (\bm r_o+\Delta \bm r)} \ud S_n \\
\nonumber &=  - e^{-jk (\hat{\bm k} \cdot \Delta \bm r - \bm {\hat r}(\theta,\phi) \cdot \Delta \bm r)} jk \frac{e^{-jkr}}{4\pi r} \bm {\hat r} \times \\
\nonumber & \qquad \int_{S_n} \bm J_e e^{jk \bm {\hat r}(\theta,\phi) \cdot \bm r_o} \ud S_n  \\
&= e^{-jk (\hat{\bm k} - \bm {\hat r}(\theta,\phi)) \cdot \Delta \bm r} \bm H^f_n(\theta,\phi),
\end{align}
where the superscript $t$ indicates that it is the translated response, and $\bm J _e$ is still the equivalent surface current of the untranslated structure. This result simply describes the angular dependent change in phase to the contribution that is seen in an observation point when moving the structure around in a lit domain. 

\subsection{Irradiance of translated structures} \label{sec:irradiance}
In this paper we will focus on one basic structure that will exist in several translated instances in the domain (that is, the structure itself will not be pertubed). This structure will be referred to as the \emph{unit structure}. Due to superposition there can in principle be several unit structures as long as they do not overlap. 

If we assume that we can find the far field radiance of this structure -- or at least an adequate approximation of that -- we can then sum over this response with the correct translations to find how a system of these structures will reflect light. The caveat in this assumption is that the structures can be small and placed so close that they couple strongly with each other, or shadow for each other, or in any other way obstruct the simple response of the structure itself. For many problems this need not be an issue: In the design of reflectarray antennas the coupling between neighboring elements have been taken into account by applying periodic boundary conditions to the single element simulation and by that obtaining adequate results for simulations, \cite{Zhou2011}; and in the analysis of the Morpho butterfly's wing's nanostructure's interaction with light it has been found that a radiation boundary conditions is sufficient, \cite{Okada2012}. 

This makes it possible to define the far-field response of this unit structure as $\bm H_0 ^f$, and we can then obtain a new expressions for the irradiance as
\begin{align}
\nonumber E  &= \half \eta \left| \sum _n e^{-jk (\hat{\bm k}- \bm {\hat r}) \cdot \Delta \bm r_n} \bm H^f_0 \right|^2 \\
&= \half \eta \left|\bm H^f_0 \right|^2 \bigg| \underbrace{\sum _n e^{-jk (\hat{\bm k} - \bm {\hat r}) \cdot \Delta \bm r_n}}_{\text{$=$AF}}  \bigg|^2.
\end{align}
In this formulation the response of the unit structure is isolated such that the total irradiance is just the response of the unit structure multiplied with the magnitude of some function squared. This function we denote the Array Factor (AF), since it plays the same role as an AF does in antenna theory, \cite{Balanis2012}. Note how the exponent inside the array factor has an angular dependent term ($\hat {\bm r}(\phi,\theta)$) and a term dependent on the direction of the incoming wave ($\hat {\bm k}$).

\subsection{Radiance of translated structures} \label{sec:radiance}
The response of the eye is not proportional to irradiance, but to radiance \cite{Dutre2006}, meaning that for appearance and color purposes, we need to convert irradiance to radiance. 

From \cite{Harvey1999} it is shown that the irradiance only having a component normal to the observation surface (which is true for a detector in the far field) is related to the radiant intensity, $I$, by
\begin{align}
I(\theta,\phi) &= r^2 E (\theta,\phi),
\end{align}
and also from \cite{Harvey1999} the relation between radiant intensity and radiance, $L$, is given as
\begin{align}
L(\theta,\phi) &= \frac{I (\theta,\phi) }{A_s \cos \theta }  = \frac{r^2}{A_s \cos \theta} E (\theta,\phi),
\end{align}
where $A_s$ is the area of the surface of the lit structure. This means that we can write the total radiance for a lit surface with repeated structures as
\begin{align}
 \nonumber L(\theta,\phi) &= \frac{r^2 \eta  }{2 A_s \cos \theta} |\bm H^f_0(\theta,\phi)|^2 \times \\
 \nonumber &\qquad \left| \sum _n e^{-jk (\hat{\bm k} - \bm {\hat r}(\theta,\phi)) \cdot \Delta \bm r_n}  \right|^2 \\
\nonumber  &= \underbrace{\frac{r^2 \eta  }{2 A_0 \cos \theta} |\bm H^f_0(\theta,\phi)|^2}_{\text{$=$unit response}} \, N \times \\
  & \qquad \Bigg| \underbrace{\inv{N} \sum _n e^{-jk (\hat{\bm k} - \bm {\hat r}(\theta,\phi)) \cdot \Delta \bm r_n}}_{\text{$=$SAF}}  \Bigg|^2, \label{eqn:fundamental}
\end{align}
where $A_0$ is the surface of $\bm H^f_0$ and $N$ is the number of summations in the sum -- which means that $A_s=N \cdot A_0$. The peculiarity of having $N$ present twice is to scale the last product such that it peaks at unity. In case of $N \rightarrow \infty$, the limit value has to be found. Since the array factor is now scaled by the number of units, we will refer to it as the Scaled Array Factor (SAF).

The expression in \eqref{eqn:fundamental} will be the foundation for all following analyses, as it shows how a prediction of the reflection from the unit structure (this we denote \emph{unit response}) and a knowledge of the position of its copies (the $\Delta \bm r _n$'s) can give a complete description of the reflected radiance. Furthermore it decouples the positioning of the structures from the response of the structure, which means that when designing e.g. the angular pattern of a given structure it is not necessary to take the shape of the structure into account and vice versa (as long as the assumptions are not violated).

\subsection{Limitations due to assumptions}
In Section \ref{sec:irradiance} and \ref{sec:radiance} we have assumed that all structures on a surface have the same response (except for the translation part), and it is important to clarify when these assumptions can be expected to hold and where one should be cautious. There are two basic problems to be cautious about: (1) the geometry surrounding the element and (2) if the surface for the far-field transformation contains the needed response of the structure.

\subsubsection{Dependence of surrounding geometry}
If an element is simulated sitting in a periodic structure like in Figure \ref{fig:structures}(a), and then in reality is sitting in some random structure like in Figure \ref{fig:structures}(b) it is clear that the actual far-field response is somewhat different, since the electromagnetic coupling to the neighboring elements has changed. It is therefore necessary that the resulting change in far-field response is small or averages out over many elements, and  it is also necessary that the displaced unit structures do not shadow each other. This will in general become less important for large unit structures, since the interelemental coupling in most cases will be negligible.

It should also be noted that all practical structures have finite sizes, and the unit structures at the edges probably will have another far-field response due to the difference in the surrounding geometry. If the surface is large compared to the area occupied by the outer unit structures, this effect should be negligible.

\subsubsection{Equivalent surface assumption}
We have assumed that there exists a plane surface, $S_0$, above the unit structure on which we can calculate the equivalent surface current, $\bm J _e$, and then find the needed unit response from here. Furthermore, we assume that we can calculate the total response from some configuration of unit structures by stitching a plane surface, $S$, together by these surface currents with a first order phase correction term taking their translation into account. 

For the above assumptions to work well we recommend to put the far-field transformation surface as close to the electromagnetic structure as possible for two reasons: the first is to take as much energy as possible into account and thereby catching the behavior the best way possible; and the second is that it will minimize unwanted contributions from the surroundings through the surface (e.g. if there are periodic boundary conditions) so that only coupling is taken into account -- and not some of the far-field from structures which in reality are not translated.

\subsection{1-dimensional version of the formulas}
In the rest of this paper, we will consider structures only with variation in the $x$- and $z$-directions and lit by waves directed in the $x,z$-plane to keep the examples simple. This means that the results will be invariant in the $y$-plane, and we therefore put $\phi=90^\circ$, \cite{Antonakakis2012}. To indicate this, we will apply the following notation in the rest of this paper:
\begin{align}
\nonumber L(\theta) &=  \frac{r \eta  }{2 d_0 \cos \theta} |\bm H^f_0(\theta,\phi=90^\circ)|^2 N \times \\
&\qquad \left| \inv{N} \sum _n e^{-jk (\hat{\bm k} - \bm {\hat r}(\theta,\phi=90^\circ)) \cdot \Delta \bm r_n}  \right|^2, \label{eqn:fundamental_1D}
\end{align}
where $d_0$ is the length of the unit structure instead of the area.

\subsection{Interpretation of the array factor}
\begin{figure*}
  \begin{tabular}{ccccc}
   Unit response: && SAF: && Total response:  \\[0.2cm]
\multirow{2}{*}[-0.65cm]{  \includegraphics[width=0.25\textwidth]{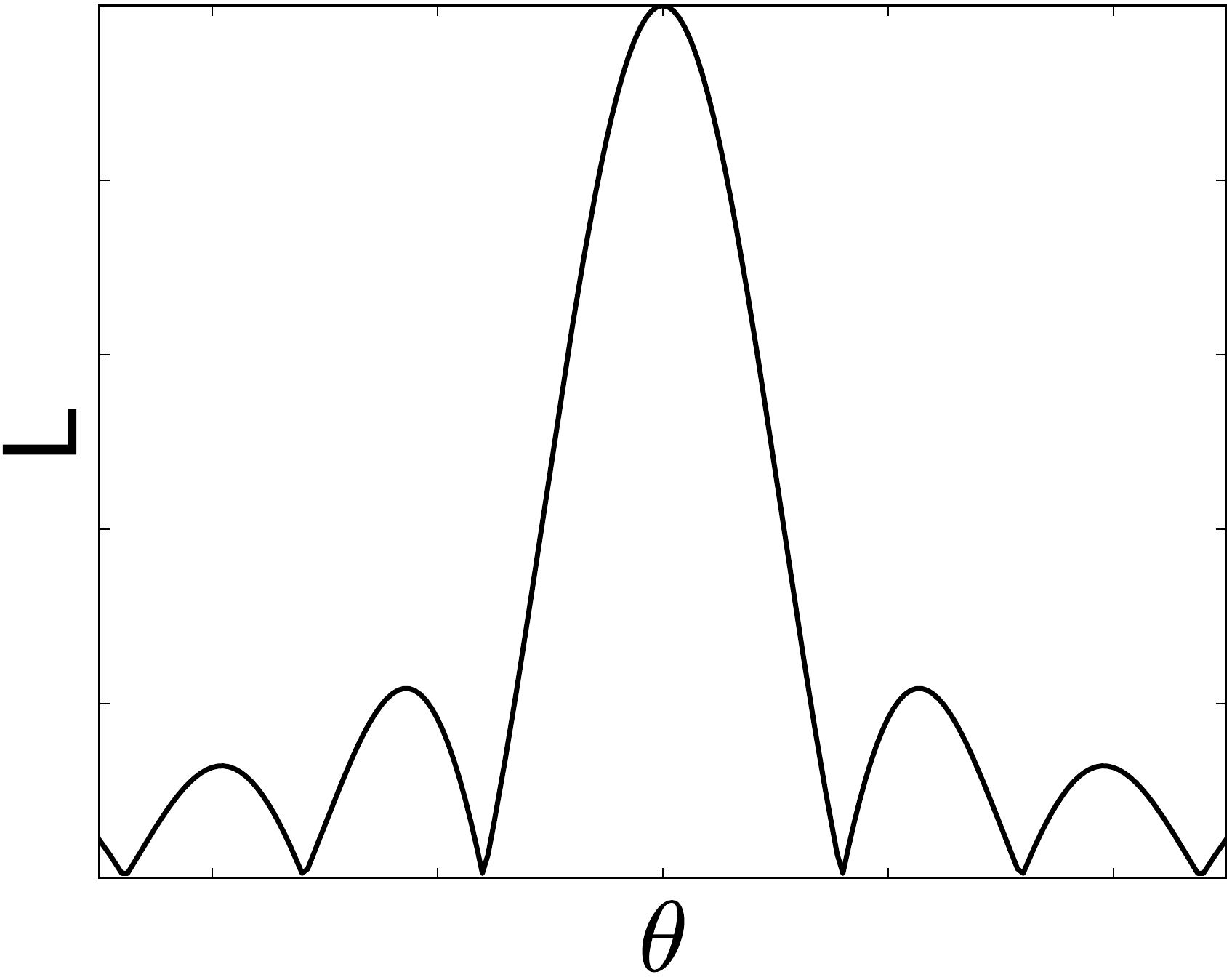}} & \Large{$ \cdot$} & 
   \raisebox{-.5\height}{\includegraphics[width=0.25\textwidth]{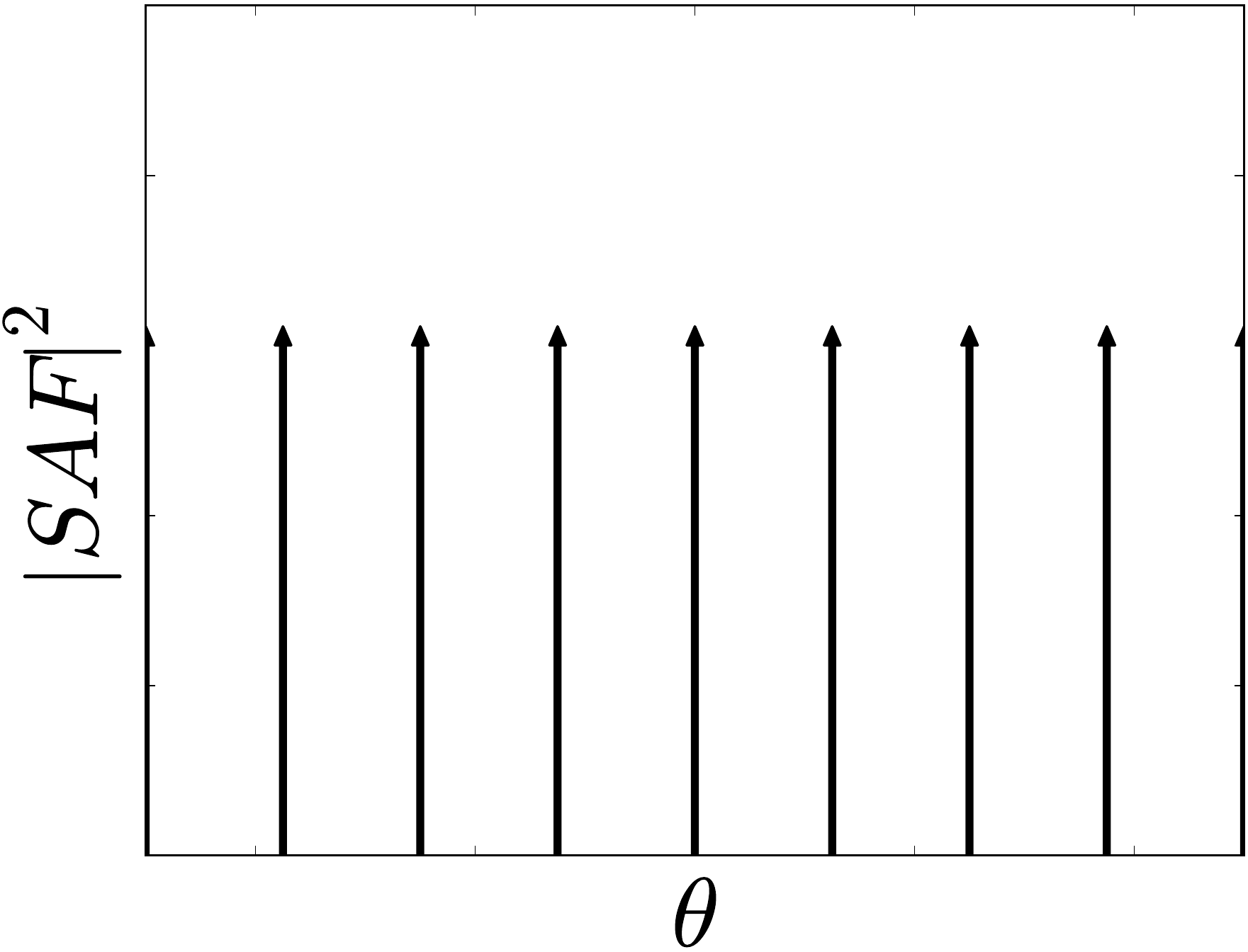}} & \Large{$=$} &
    \raisebox{-.5\height}{\includegraphics[width=0.25\textwidth]{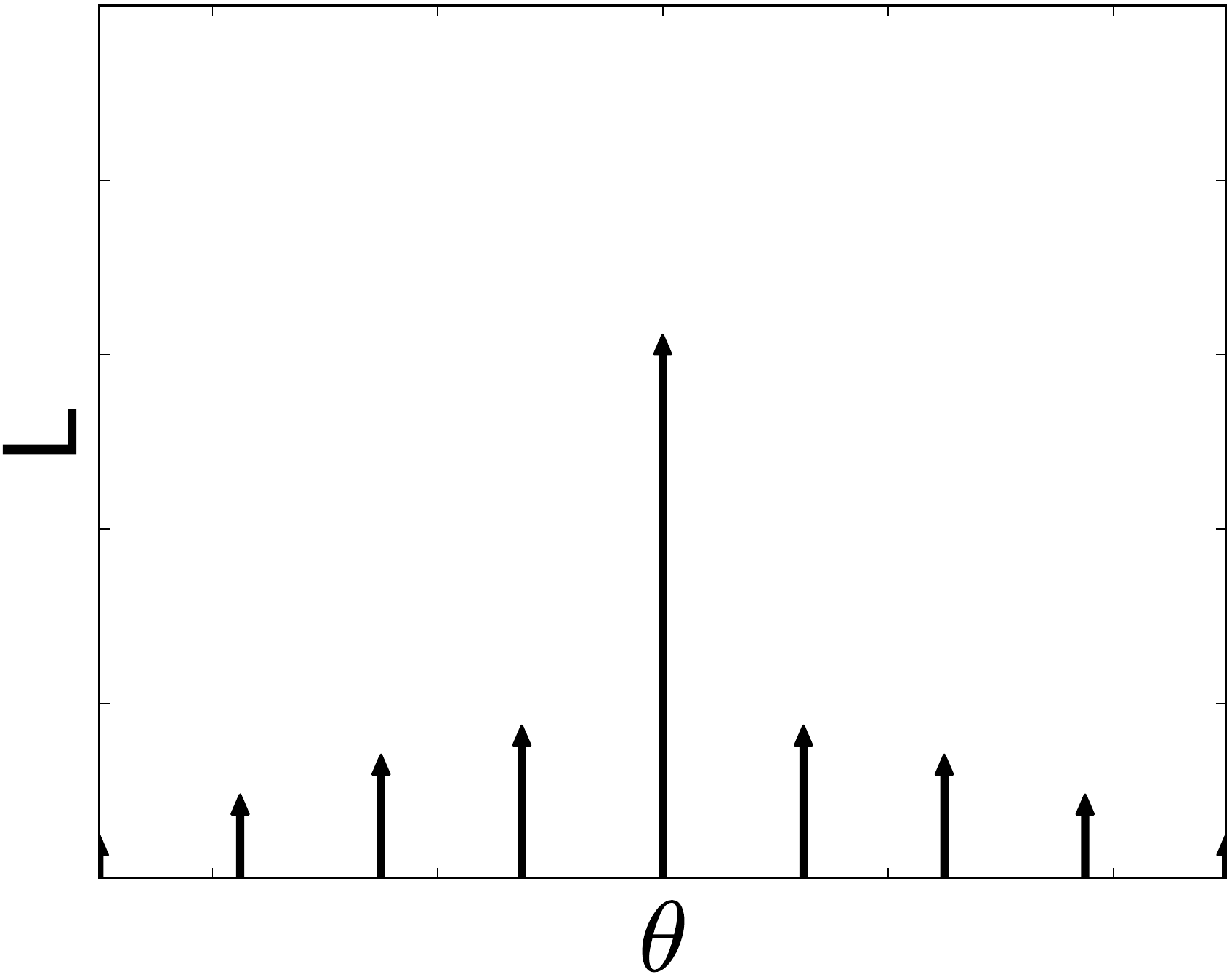}} \\
     &  & \texttt{ }(b) & & \texttt{ }(c) \\[0.2cm]
\multirow{2}{*}[0.15cm]{(a)}   & \Large{$ \cdot$} & 
   \raisebox{-.5\height}{\includegraphics[width=0.25\textwidth]{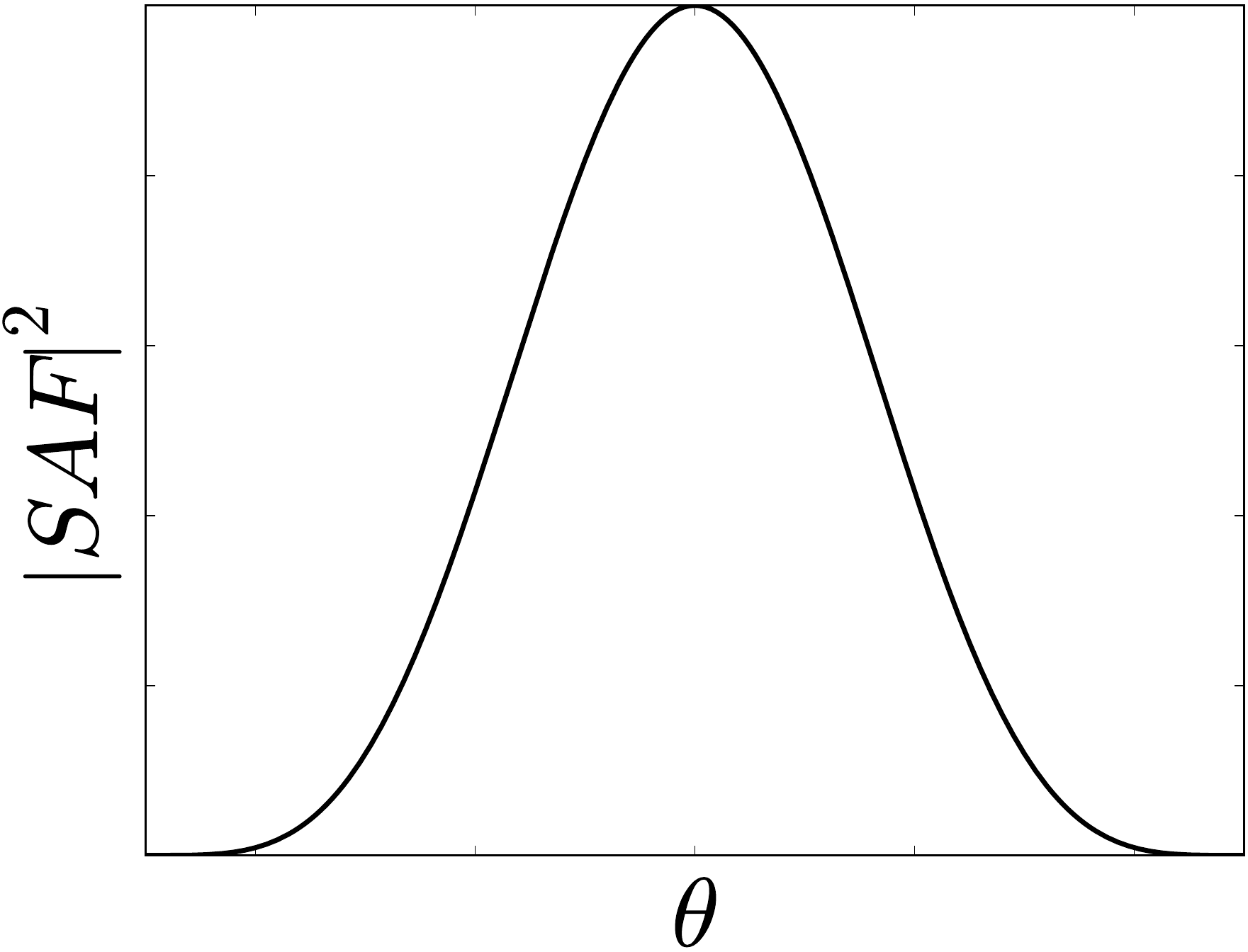}} & \Large{$=$} & 
   \raisebox{-.5\height}{\includegraphics[width=0.25\textwidth]{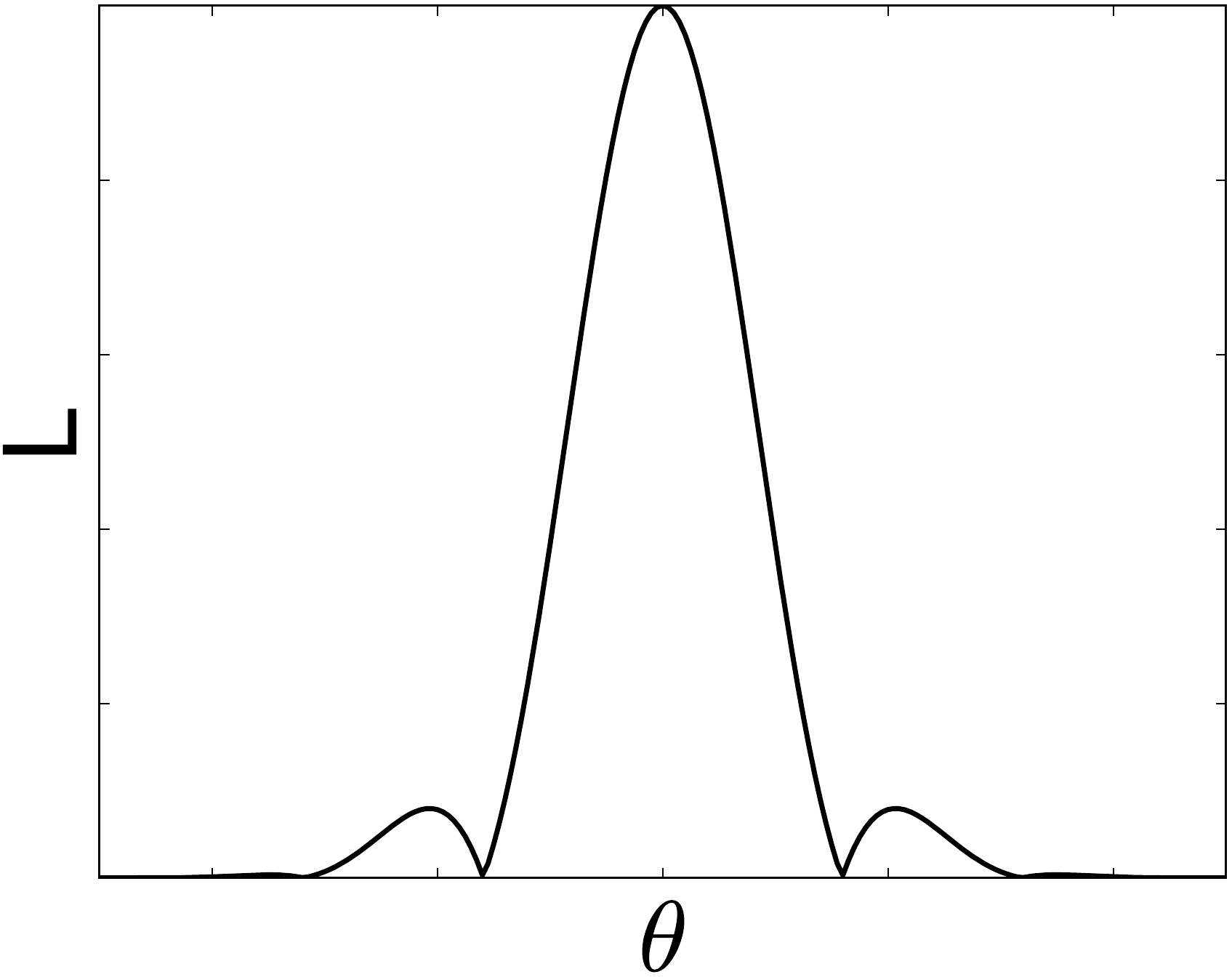}} \\
        & & \texttt{ }(d) & & \texttt{ }(e)
  \end{tabular}
   \caption{Example of how a structure with some unit response (a) is influenced by different SAF's. In (b) the SAF for a strictly periodic structure is shown (the comb function), and the effect of combining the structure giving the unit response in (a) with the pattern giving rise to the SAF in (b) is shown in (c). The arrows indicate that all energy is emitted at discrete points. If the SAF is like in Figure (d), then the response will be as seen in (e).}
   \label{fig:AF}
 \end{figure*}

From \eqref{eqn:fundamental} and \eqref{eqn:fundamental_1D} it is seen how the total radiance is the product of a unit radiance and the absoulte squared SAF. This means that they can be treated independently of each other, and in this paper we will focus only on the SAF and not consider a specific unit response, but instead pose the problem: \emph{ if we have designed a structure, in which way should it be distributed to give the reflection we want?} The effect of different choices of translations (that is, different SAF's) of the same structure and its influence is exemplified in Figure \ref{fig:AF}.

\section{Analysis of random structures}\label{sec:analysis}
This section will apply the previous obtained framework to different cases -- all focusing on explaining the influence of (random) translations of well-defined structures as observed by experiments and simulations in the literature. Besides the obvious aim of giving a mathematical tool to analyze random translations, these examples are also chosen to give the reader an intuitive understanding of the mechanisms behind the obtained effects.

As stated earlier, all results will be in 2D and furthermore with an incoming wave parallel to the $z$-axis. This means that some important features like the dependence on the direction of the incoming wave, $\hat{\bm k}$, will not be discussed in this paper.

\subsection{No randomness -- a special case called diffraction}
To introduce the concept and confirm its validity for a known (but trivial) case, the first example will be the rather simple case of a repeated structure with no randomization.

\begin{figure}
 \centering
 \includegraphics[width=\columnwidth]{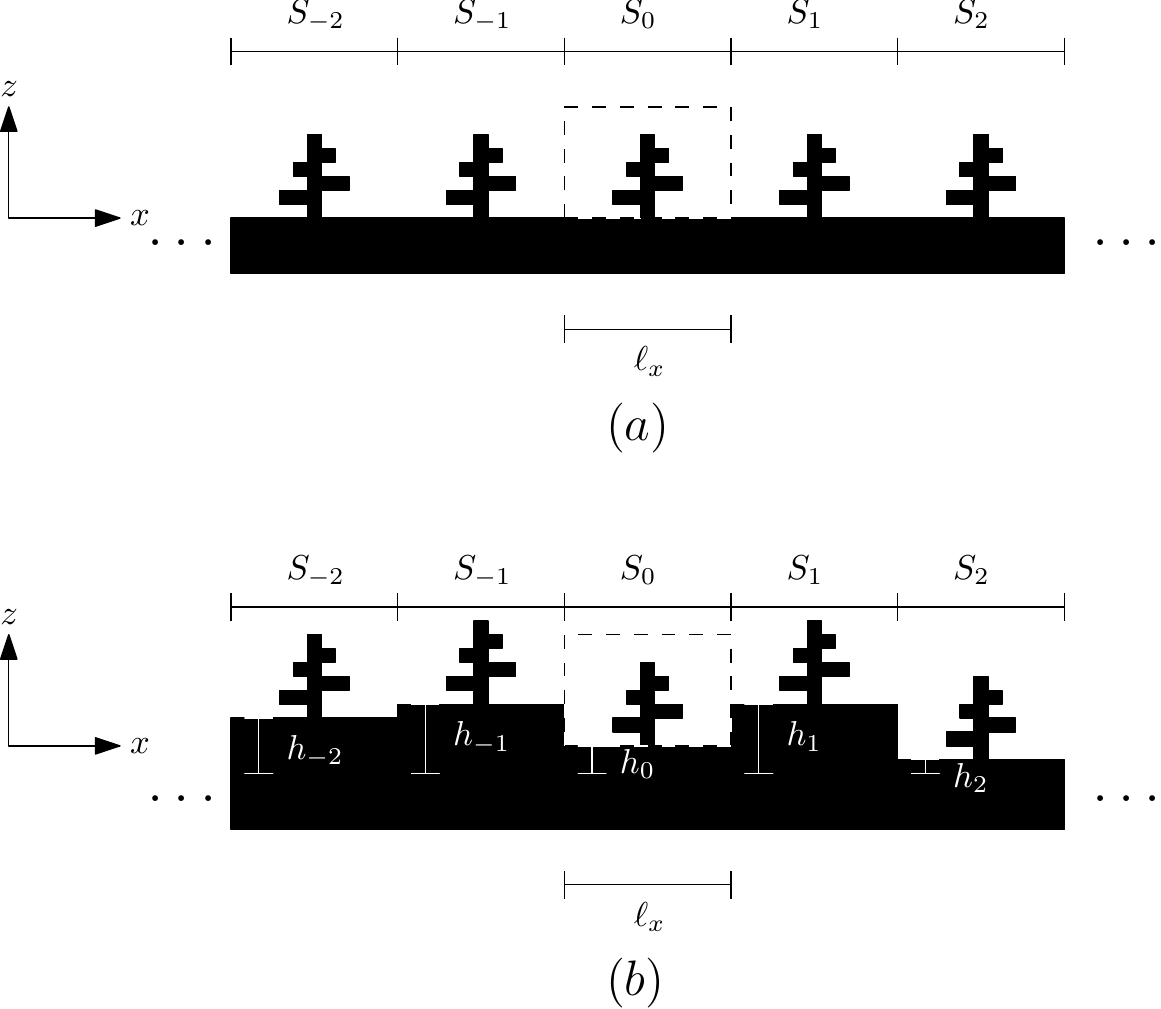}
 \caption{(a) Strictly periodic structure, (b) The same structure but with per period height translations.}
 \label{fig:structures}
\end{figure}

Consider an inifinitely periodically repeated unit structure within a domain, see Figure \ref{fig:structures}(a). If this structure is lit with a normal incidenct plane wave, then we have that

\begin{align}
 \hat{\bm k} = (0,0,-1), \quad \Delta \bm r_n = (n \ell _x, 0, 0) \, , n \in \Z ,
\end{align}
where $\ell_x$ is the period with which the structure is translated. The scaled array factor for this structure then becomes
\begin{align}
\nonumber  SAF(\theta) = \inv{N} \sum _{n \in \Z} e^{-jk (\hat{\bm k} - \bm {\hat r}(\theta)) \cdot \Delta \bm r_n}  \\
	    = \inv{N} \sum _{n \in \Z} e^{jk n \ell_x \sin \theta }  .
\end{align}
For infinite repetition, the limit of the summation inside the norm can be expressed as, \cite{Bracewell2000},
\begin{align}
\nonumber \sum _{n \in \Z} e^{j2\pi  n \frac{\ell_x}{\lambda} \sin \theta}  &= \sum _{n\in Z} \delta \left(\frac{\ell_x}{\lambda} \sin \theta-n \right)  \\
& =: \comb \left( \frac{\ell_x}{\lambda} \sin \theta \right),
\end{align}
where $\delta$ is the Dirac delta function, and the name \emph{comb} is given due to the function's resemblance with a comb. It is also sometimes referred to as the \emph{shah} function. The comb function can be seen in figure \ref{fig:AF}(b). Since $\delta (x) = 0, \, \forall x \in \R \backslash \{0\}$, reflection from this structure will only appear when 
\begin{align}
\frac{\ell_x}{\lambda} \sin \theta = m \in \Z \label{eqn:grating_equation}
\end{align}
and have quite strong intensity. The relation in \eqref{eqn:grating_equation} is called the \emph{grating equation} and $m$ is normally referred to as the \emph{mode number}. It is a well-known result, explaining e.g. the rainbow effect seen when observing a Compact Disc (CD), which consist of equally spaced grooves that are used to store the data. This behavior is loosely referred to as diffraction, and the framework is seen to explain this behavior as expected.

\subsection{Random height variation of repeated structure}
Consider the same setup as before with an infinitely repeated structure, see Figure \ref{fig:structures}(a). In the analysis of the coloration of the Morpho butterfly, numerical simulations in \cite{Lee2009} shows that by adding a random height variation to each unit, the total response will start to resemble the unit response with overlayed high frequency ripples. To analyze this we apply a random height translation to each unit drawn from a uniform distribution with values between 0 and $\lambda_{max}$, where $\lambda_{max}$ is the longest wavelength in the analysis. The final structure will then look like the one seen in Figure \ref{fig:structures}(b). The incoming wave is still normal to the surface, so the components for the SAF of this system are
\begin{align}
 \hat{\bm k} = (0,0,-1), \quad \Delta \bm r_n = (n \ell_x, 0, \ell_{z,n}) \, , n \in \Z,
\end{align}
where $\ell_x$ is the period and $\ell_{z,0},\ell_{z,1},\ell_{z,-1},\ell_{z,2} \ldots$ is a sequence of numbers drawn from a uniform distribution with values between 0 and $\lambda_{max}$. This mean that the AF now takes the form
\begin{align}
\nonumber SAF(\theta) &= \inv{N} \sum _{n \in \Z} e^{-jk (n \ell_x\sin \theta  - (\cos \theta+1) \ell_{z,n})}    \\
 &=  \inv{N} \sum _{n \in \Z} e^{-jk  n \ell_x \sin \theta} e^{ jk (\cos \theta +1) \ell_{z,n}}  , \label{eqn:fullrandom}
\end{align}
where we see that the first product in the summation comes from the periodic translation, and the second from the random height translation. The minus sign in the second product indicates that positive height displacements reflect the phase earlier, and the angle dependent $1+\cos \theta$ can be interpreted geometrically as the extra added distance the wave has to travel. That is, for specular reflection where $\theta=0$, the wave will also travel the same phase \emph{less} than it did when reaching the structure, but for other values of $\theta$, it will travel a bit longer, and therefore not as much negative phase lag will be removed.

\begin{figure*}[!htb]
\centering
\begin{tabular}{cc}
 \includegraphics[width=\columnwidth]{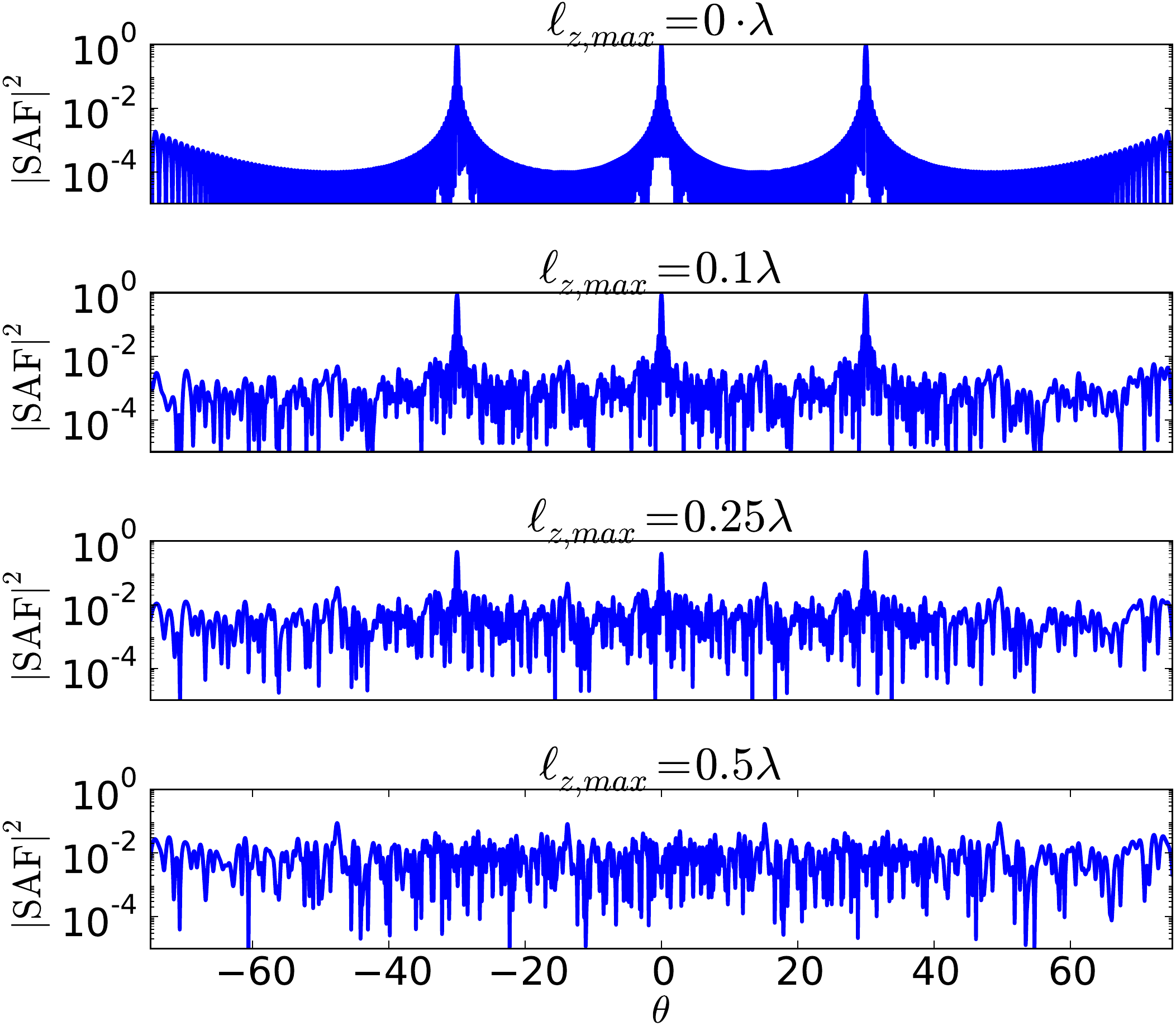} &
 \includegraphics[width=\columnwidth]{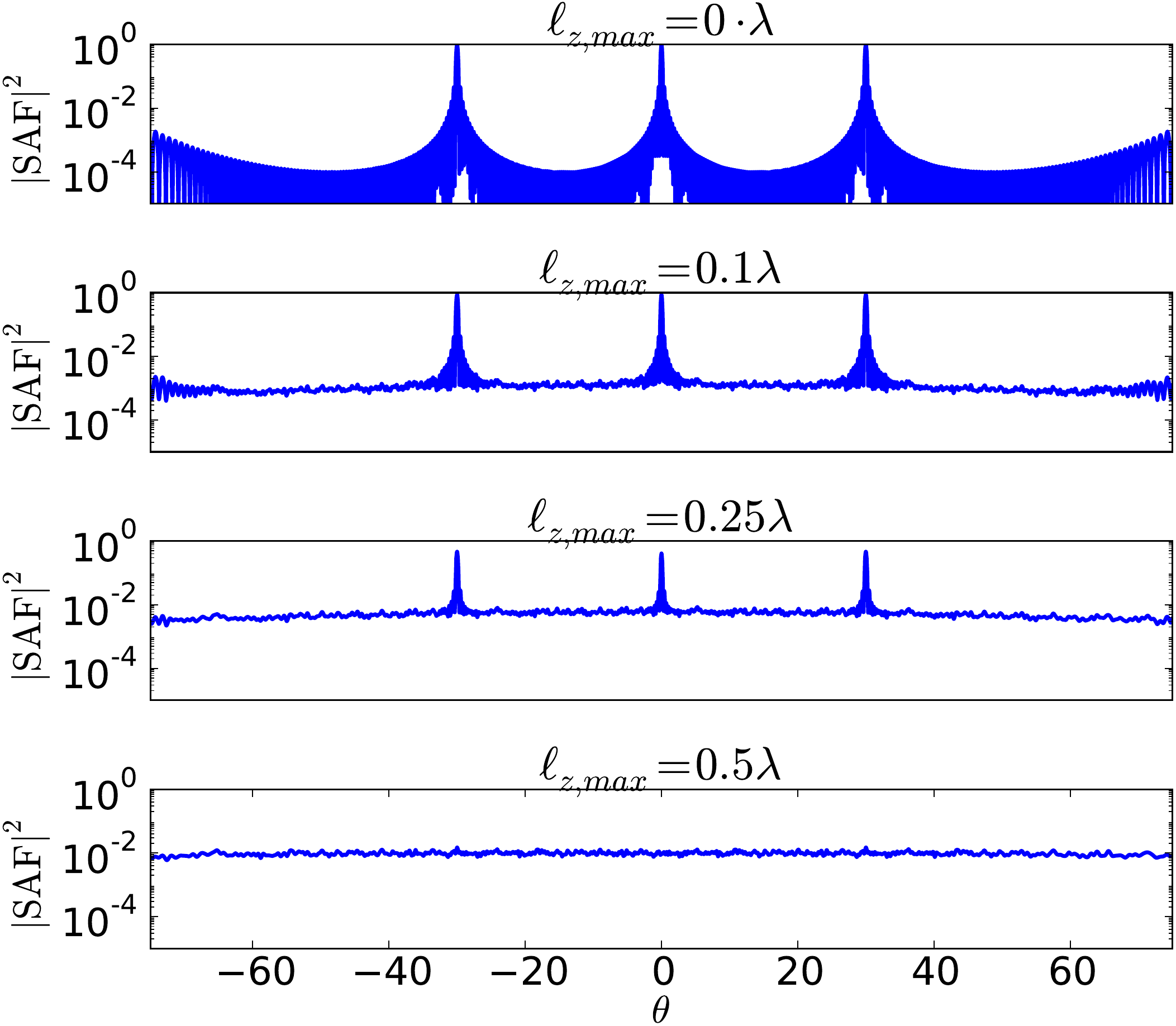} \\
   (a) & (b) 
\end{tabular}
 \caption{(a) Numerical calculation of \eqref{eqn:fullrandom} with $\ell_{z,n}$ picked uniformly from $[0, \ell_{z,max}]$ and $N=100$ repetitions and a periodicity in $x$ of $\ell_x = 2\lambda$. By changing $\ell_x$ the SAF would either be dilated or constricted such that the grating modes still match \eqref{eqn:grating_equation}. (b) 100 averages of the setup in (a) using \eqref{eqn:incoherentfullrandom}.}
 \label{fig:fullrandom} 
\end{figure*}

A typical response of \eqref{eqn:fullrandom} is seen in Figure \ref{fig:fullrandom}(a) for a finite number of structures ($N=100$). It it seen how it is required that $\ell_{z,n}$ should vary between zero and half a wavelength to get a SAF where no diffraction pattern is present and also that if there is no randomization then the SAF has sharp intensities peaking in the grating modes that can be calculated from \eqref{eqn:grating_equation} -- the finite number of repetitions spreads out the intensities from just a single angle to a small angular area. It is also seen that there is a gradual change from a diffraction to no diffraction with the gradual variation of wavelength. For larger maximum values of $\ell_{z,n}$, the diffraction is in general not strongly present.

A good way of describing the above observations intuitively is by considering what phase is most probable to be observed at a certain point, and if there is no preferred phase then there is no possibility of interference, whereas if there is a preference of a phase, then that will give rise to interference effects (e.g. if the phase is varying in a smaller interval than $\theta \in [0^\circ,360^\circ]$).

These results can be confirmed by \cite{Saito2011} where  they report a clearly visible diffraction pattern for small variation and show that larger variations are needed to minimize the effect. Furthermore it can be seen in Figure \{4(c),\cite{Saito2011}\} in \cite{Saito2011} how the $380\unit{nm}$ still shows the first order diffraction at $\theta =  \arcsin \frac{380}{400} \approx 72^\circ$, and otherwise a random pattern with strong ``noise'' and a zero order mode still present since the variations are smaller than a wavelength, and furthermore how longer wavelengths -- as just shown -- gives a larger contribution to mode 0 (when the relative reflection from Figure \{3,\cite{Saito2011}\} is taken into account). Furthermore, the same analysis reveals the governing effect of Figure \{6,\cite{Steindorfer2012}\} in the explanation of the results in \cite{Steindorfer2012}.

\subsubsection{Added incoherence}
In reality, the randomized spectrum has less ripples than in Figure \ref{fig:fullrandom}(a). In \cite{Kinoshita2002,Saito2011} this has been taken into account by averaging the irradiance of a larger number of samples to add the effect of incoherence. The explanation for doing so is that the phases of two uncorrelated waves on average will neither add destructively or constructively and it is therefore possible just to sum their powers.  That is,
\begin{align}
|SAF(\theta)|^2 &= \left< \left| \inv{N} \sum _{n \in \Z} e^{-jk  n \ell_x \sin \theta} e^{ jk (\cos \theta +1) \ell_{z,n}}  \right|^2 \right>,\label{eqn:incoherentfullrandom}
\end{align}
where $\left< \cdot\right>$ for this equation indicates that the average will be taken of the SAF for many different seeds of $\ell_{z,n}$. By doing this averaging we end up with the result in Figure \ref{fig:fullrandom}(b), where the ``noise ripples'' are now much smaller. This is in good agreement with the results from \cite{Saito2011}, and for an even higher number of averages, the ripples become even smaller.

\subsection{Triangular height distribution}
\begin{figure}
\centering
\begin{tabular}{c}
 \includegraphics[width=\columnwidth]{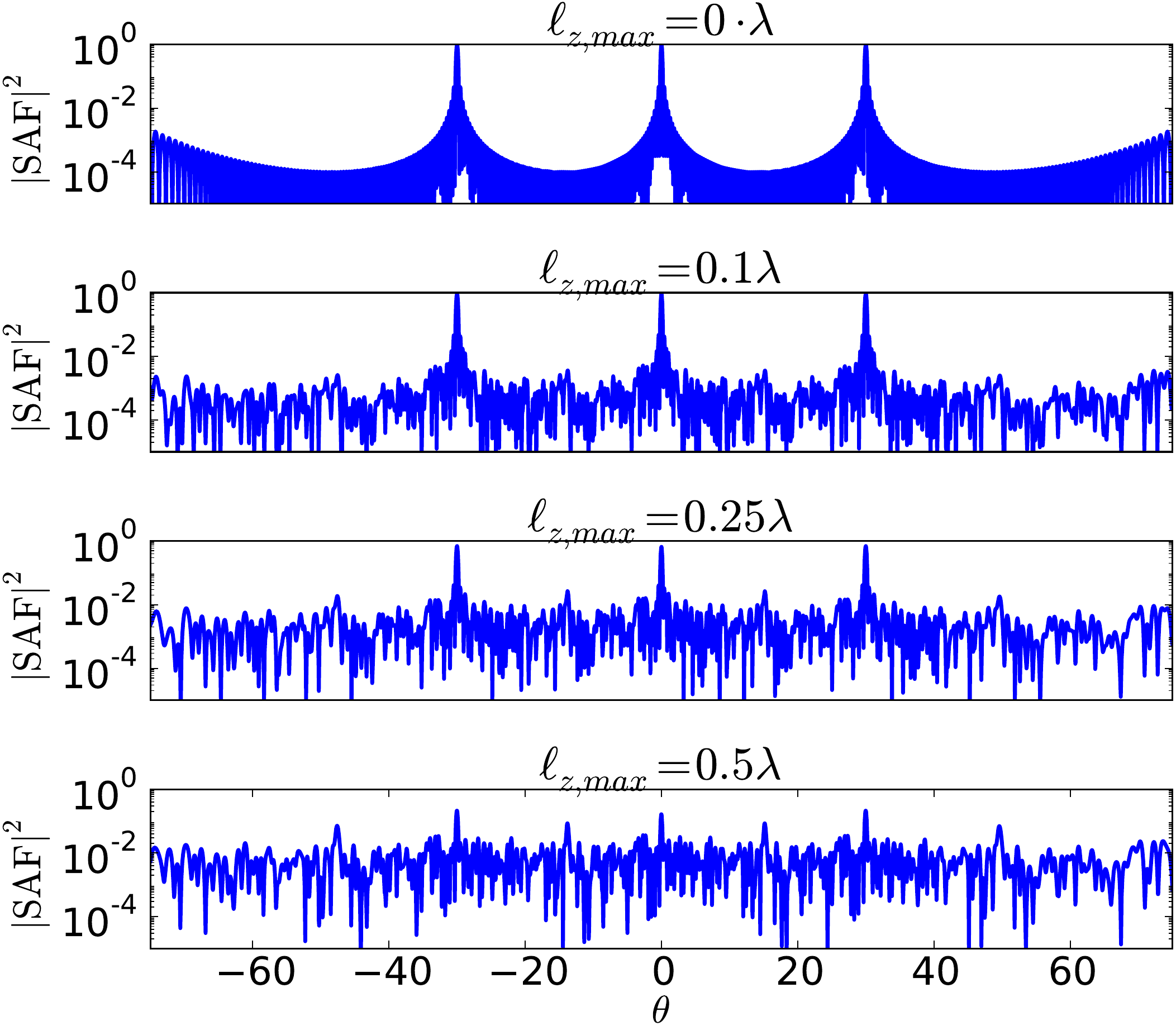}
\end{tabular}
 \caption{Results for the same setup as in Figure \ref{fig:fullrandom}(a), but now with $\ell_{z,n}$ taken from a \emph{triangular} distribution.}
 \label{fig:triangularrandom} 
\end{figure}

\begin{figure}
\centering
 \includegraphics[width=\columnwidth]{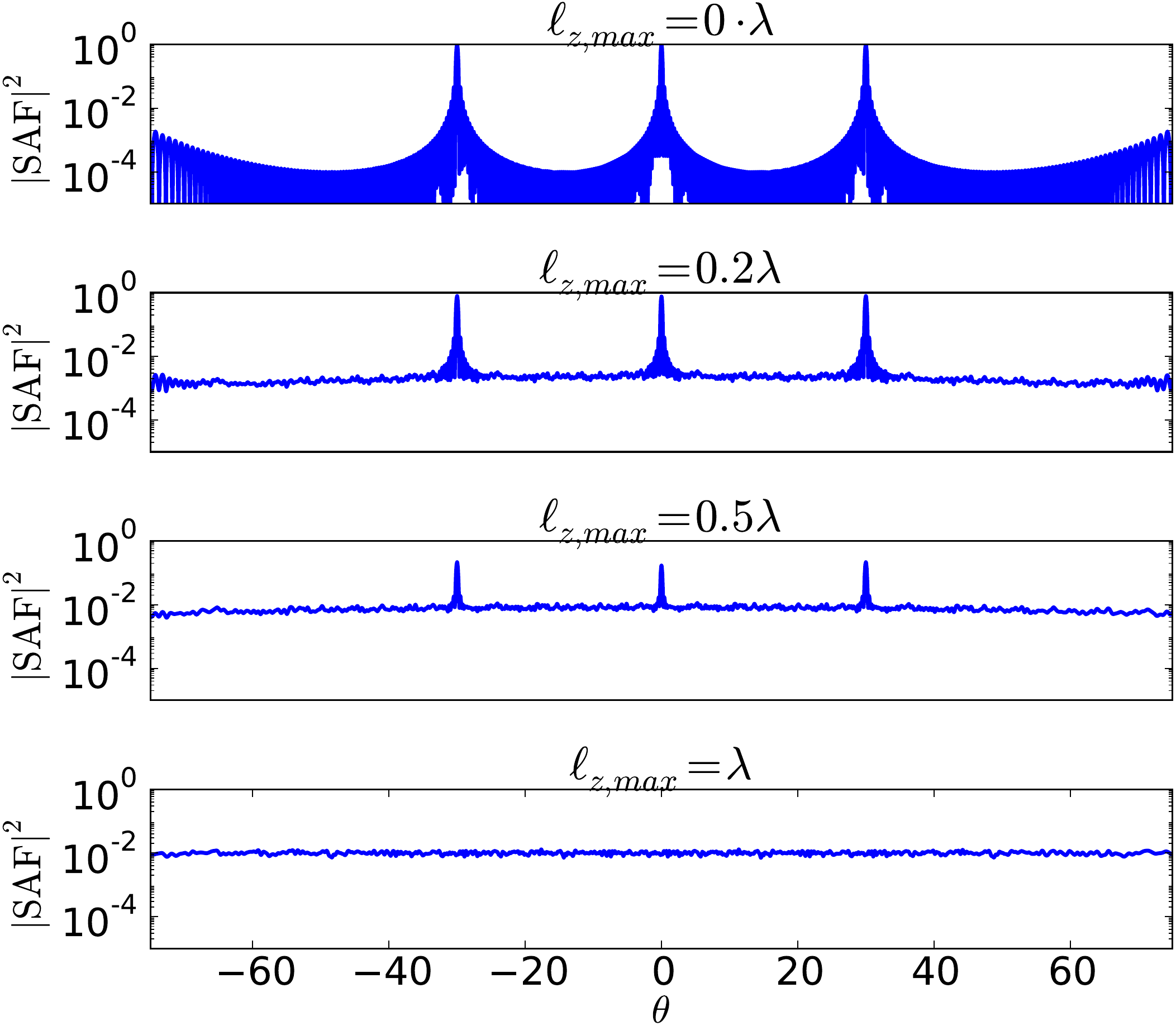}

 \caption{100 averages of the same setup as \ref{fig:triangularrandom}(a) -- but for \emph{double} the interval width of the distribution used in \ref{fig:triangularrandom}(a)!}
 \label{fig:triangularrandomdouble} 
\end{figure}

In \cite{Lee2009} the height displacements are chosen from a triangular distribution. Using \eqref{eqn:fullrandom} where $\ell_{z,n}$ is now drawn from a triangular distribution ranging from 0 to $\ell_{z,max}$ with the triangular peak placed in $\ell_{z,max}/2$ gives the results presented in Figure \ref{fig:triangularrandom}. This result has more visible diffraction than for the uniform samples. In particular it is seen how they are still present for $\ell_{z,max}=1/2\lambda$. Investigating the Ph.D. thesis on which the article is based, \cite{Lee2009a}, reveals that the interval for the triangular distribution is also double the size of the uniform distribution it is compared with. Doing the analysis with a broader distribution gives the results in Figure \ref{fig:triangularrandomdouble}, and from here it is seen that the modes actually are better suppressed, which explains the choice of distribution in \cite{Lee2009}. We have chosen to plot the incoherent/averaged version, since it is then easier to compare it with Figure \ref{fig:fullrandom}(b).

\subsection{Binary randomized heights}
\begin{figure}
\centering
\includegraphics[width=\columnwidth]{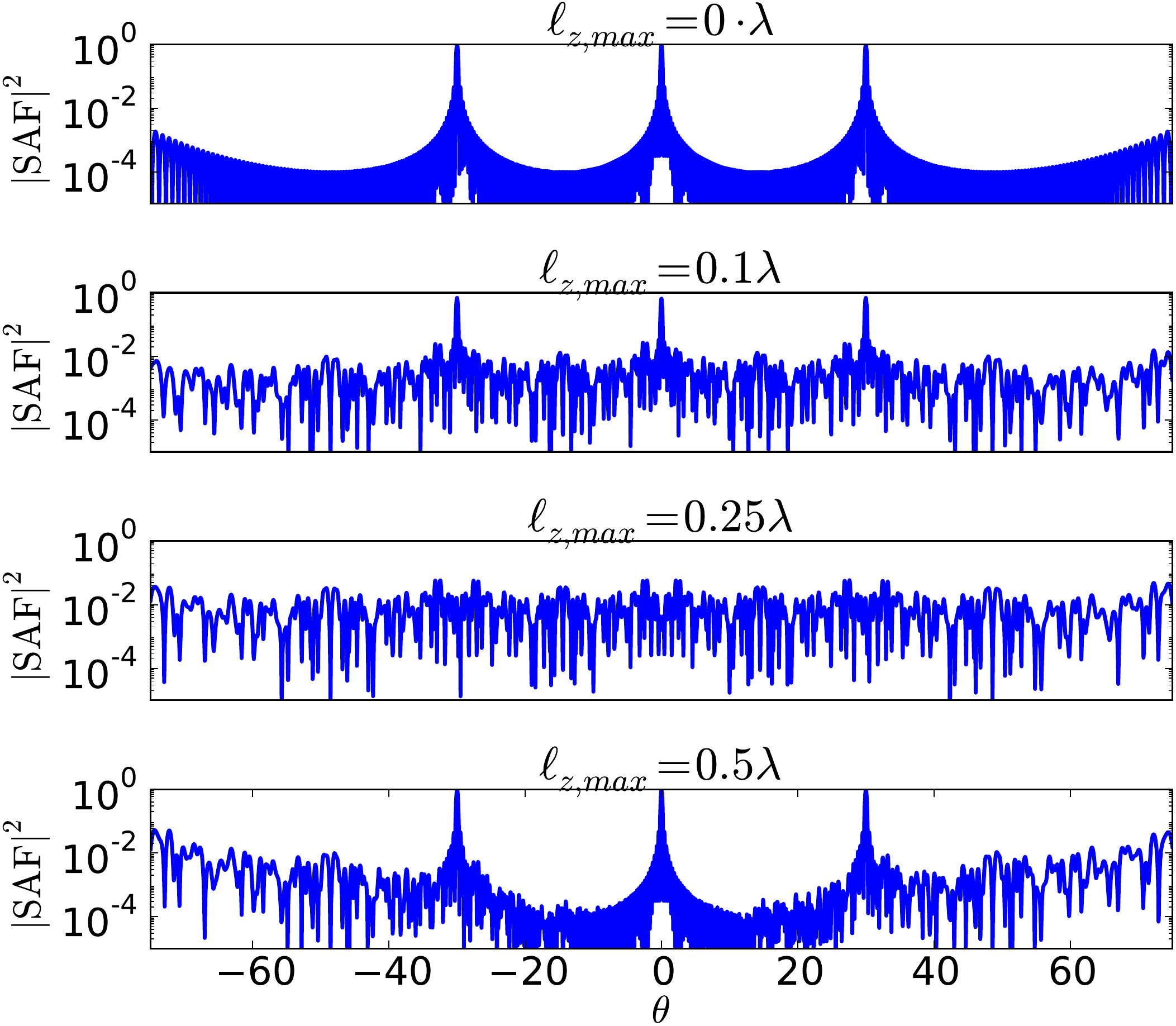}
 \caption{Results for the same setup as in Figure \ref{fig:fullrandom}(a), but now with $\ell_{z,n}$ only taking the values $0$ and $\ell_{z,max}$.} 
\label{fig:binrandom} 
\end{figure}

In \cite{Saito2009} a blue surface without specular reflections is produced to mimic the behavior of the Morpho butterfly wing. For easy fabrication purposes, the randomization of the unit structure is made up by only two heights. that is, $\ell_{z,n}$ in \eqref{eqn:fullrandom} can only take on two values: $0$ and $\ell_{z,max}$. Repeating the setup from before, but now with this binary randomness, we obtain the results presented in Figure \ref{fig:binrandom}. Here it is seen how a height difference of a quarter of a wavelength at the target frequency is needed to get a flat $|SAF|^2$. This is in perfect agreement with the height choosen in \cite{Saito2009}. Choosing a height difference on half a wavelength gives strong grating modes, though. The reason for this is that at a height difference of $1/4 \lambda$ the reflected light is sent back $180^\circ$ out of phase and therefore interferes destructively, leaving no preferred directions for the reflected light, whereas if the light travels $1/2 \lambda$ extra back and forth it corresponds to $360^\circ$, which gives constructive interference and modes like seen in Figure \ref{fig:fullrandom}(a). Since the reflected light travels a bit longer than $1/2\lambda$ when reflected back in off specular directions (taken care of in the equation by the $\cos \theta$ term), the effect of constructive interference wears off more and more for larger and larger angles, which is also seen on the plot when comparing $\ell_{z,max}=0$ with $\ell_{z,max}=0.5\lambda$.

All in all, this clearly shows the need for analyzing the random design for specific situations, as here $\ell_{z,max}$ has to be chosen differently, and also the effect here can only be expected to work well for a smaller range of wavelengths than the uniform randomness. In the design in \cite{Saito2009}, the wavelengths far from the blue region around $440 \unit{nm}$ are suppressed by the unit structure, thus hiding the diffraction effects that would otherwise have been present.

\begin{figure}
\centering
\includegraphics[width=\columnwidth]{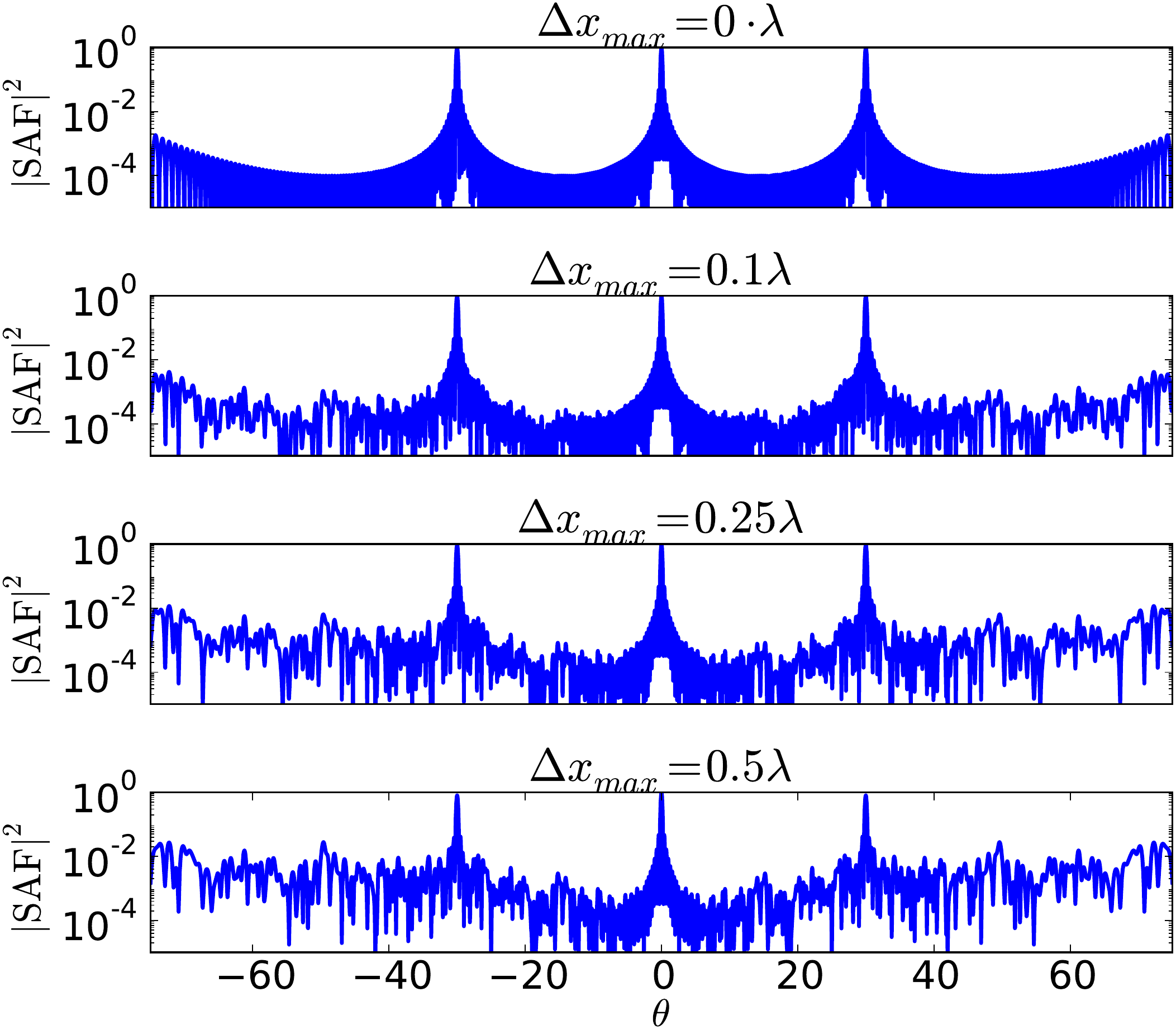}
 \caption{Results for the same setup as in Figure \ref{fig:fullrandom}(a), but with in-plane movement as specified in \eqref{eqn:inplanecomponents}.}
 \label{fig:xrandom} 
\end{figure}

\subsection{In-plane translated elements}
In \cite{Saito2011} they also consider in-plane movement -- that is, movement in the $x$-direction of a repeated structure. In doing this, we would need the following definitions for the SAF:
\begin{align}
 \hat{\bm k} = (0,0,-1), \quad \Delta \bm r_n = (n \ell_x+\Delta x_{n}, 0, 0),  \quad n \in \Z, \label{eqn:inplanecomponents}
\end{align}
where the $\Delta x_{n} $'s is a a sequence of random variables drawn uniformly from the range $[-\Delta x_{max}/2, \Delta x_{max} /2 ]$. This gives rise to the following SAF:
\begin{align}
\nonumber SAF(\theta) &= \inv{N} \sum _{n \in \Z} e^{-jk (n \ell_x+ \Delta x_n)\sin \theta  }   \\
 &= \inv{N} \sum _{n \in \Z} e^{-jk  n \ell_x \sin \theta} e^{- jk \Delta x_n\sin \theta  } . \label{eqn:xrandom}
\end{align}
By simulating \eqref{eqn:xrandom}, we get the results shown in Figure \ref{fig:xrandom}. Comparing this with \eqref{eqn:xrandom} it can be seen that in-plane translation will never affect the specular mode response (mode 0) since $\sin \theta=0$ in that direction which means that $SAF(0)=1$ no matter the randomization. For larger angles the effect will be more and more prominent, though, since $\sin \theta$ is larger and the preferred phase will be less prominent in the phase distribution. This is a huge limitation with respect to creating an effect with no visible diffraction, but could add some randomness to large angles. This is also what is observed in \cite{Saito2011}.

It should be noted that the displacements in Figure \ref{fig:xrandom} are quite large, and it would require a structure with lots of air in between as for the structure in this example not to overlap or couple significantly.

\section{Color representation of SAF's}\label{sec:colors}
\begin{figure*}
 \centering
\includegraphics[width=0.8\textwidth]{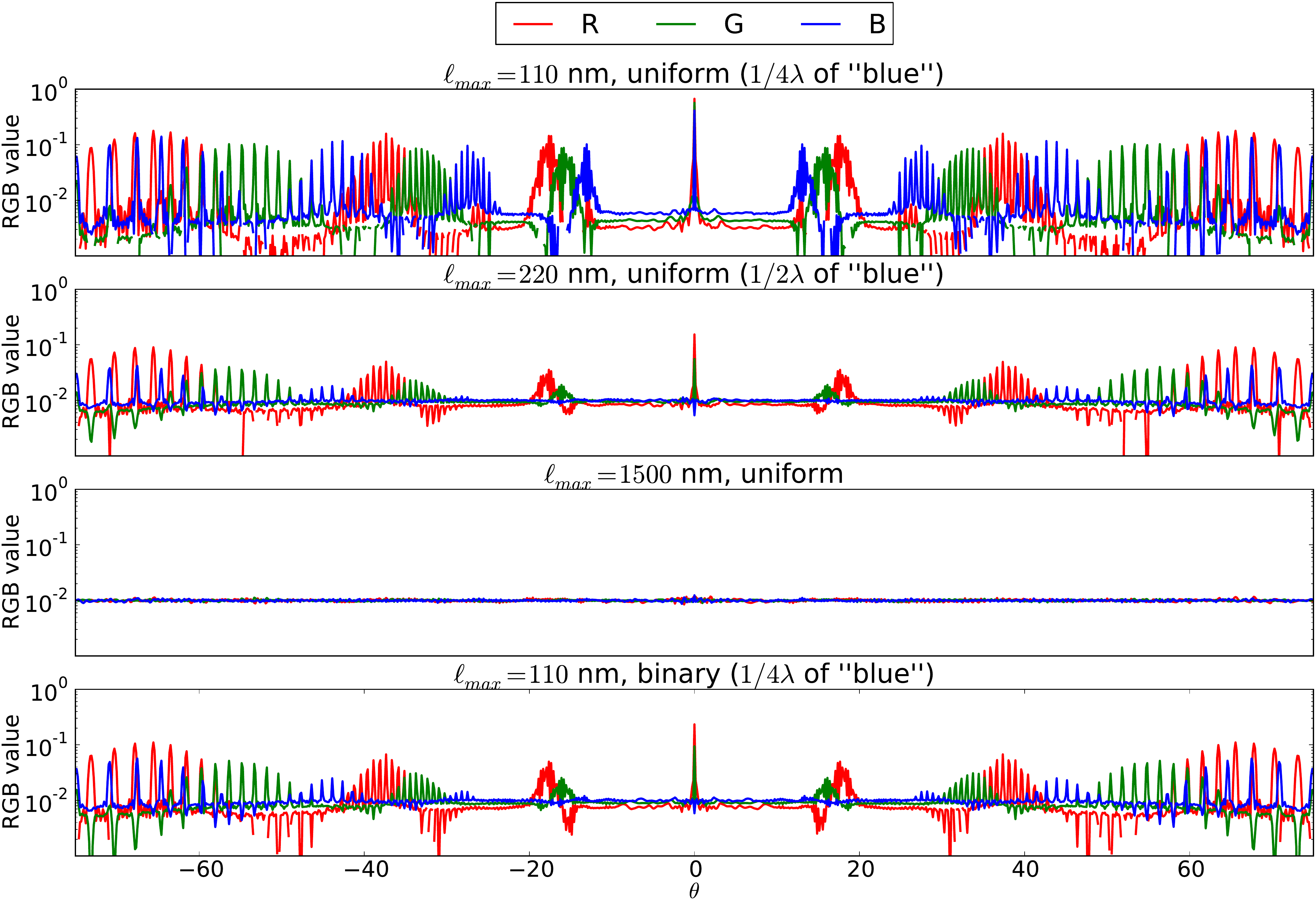}
\caption{Color representation of different SAF's. All unit structures are repeated with a period of $2 \unit{\mu m}$ and incoherence has been taken into account by averaging over 200 samples. The oscillatory effect seen at large angles is due to the fact that the angular spacing between wavelengths gets larger. Notice how the response for blue is flat for all plots except the uniform random distribution only going to $110 \unit{nm}$.}
\label{fig:color_analysis}
\end{figure*}

To give an idea of the interpretation of the obtained results in term of color effects for a surface, this section will present some of the SAF's converted to RGB-values for given sizes. The examples used all have a period of $\ell_x = 2 \unit{\mu m}$, and differ by (1) a random height variation drawn from a uniform distribution between $0$ and $110\unit{nm}$; (2) a random height variation drawn from a uniform distribution between $0$ and $220\unit{nm}$; (3) a random height variation drawn from a uniform distribution between $0$ and $1500 \unit{nm}$; (4) and a uniform binary height displacement with the values $0$ and $110 \unit{nm}$, which corresponds to the parameters chosen in \cite{Saito2009}. The results are presented in Figure \ref{fig:color_analysis}.

What can be seen from these color plots is that if a uniform random distribution is chosen, and we are designing for a color around a certain wavelength, $\lambda$, then $1/2 \lambda$ should be chosen as the upper limit for the uniform distribution (blue has a wavelength around $440 \unit{nm}$), whereas a binary random distribution will require only $1/4 \lambda$, and as shown earlier, choosing $1/2 \lambda$ will give a strong diffraction pattern. Furthermore it is seen, that chosing a large upper limit of the uniform distribution makes a good choice for giving a flat distribution of all colors. This could have some practical issues, though, since the translation may be large compared to the unit structure and possibly violate the assumptions stated earlier. All in all, this suggests that when designing random patterns for a given color, it may be beneficial to use binary randomness if possible, since this gives the smallest translation of the structures and in many cases will be easier to produce by e.g. an etching process as in \cite{Saito2009}. Not all colors can be represented by just one relatively small band in the visible range like e.g. magenta or white, and if that is not possible other means are needed. This could for example be large uniform randomness or possibly the concept of binary randomness but expanded to more levels which are chosen such that they provide destructive interference at several wavelengths. The method will be dependent on the target color, and further investigations are therefore left for the concrete cases that a scientist or a designer may have.

\section{Conclusions} \label{sec:conclusion}
For the first time, a framework for describing randomness of repeated structures has been presented and it has been shown how earlier observations in the literature all can be explained by this framework. The results presented are intented to push forward the understanding of these randomization phenomena -- for example it explains why height randomization removes diffraction patterns much better than in-plane randomization, and how to test the effect of different height distributions. This saves time and gives more insight in the analysis of random phenomena compared to the more expensive full-wave simulation of repeated structures as has been seen earlier in the investigation of random effects. Even more important, the framework makes it possible to apply a systematic approach for chosing randomization characteristics when designing surfaces with new color effects based on a unit structure. This, in turn, makes it easy to ellaborate on the produced surfaces in e.g. \cite{Saito2009} for different color effects.

\section*{Acknowledgments}
The author would like to thank Professors Olav Breinbjerg and Ole Sigmund for discussions and excellent feedback regarding this paper. The author is grateful for the support from the Danish National Technology Foundation through the ODAAS project.

\newcommand{\josaa}{J.\ Opt.\ Soc.\ Am.\ A}%
\newcommand{\josab}{J.\ Opt.\ Soc.\ Am.\ B}%
\newcommand{\ao}{Appl.\  Opt.}%
\newcommand{\opex}{Opt.\ Express}%

\end{document}